\documentclass[%
 reprint,
showpacs,preprintnumbers,
 amsmath,amssymb,
 aps,
prc,
superscriptaddress]{revtex4-1}

\usepackage{graphicx}
\usepackage{dcolumn}
\usepackage{bm}

\usepackage{hyperref}
\usepackage{booktabs}
\usepackage{subfigure}
\usepackage[pdftex]{color}

\begin{document}


\title{QCD phase diagrams via QHD and MIT based models}

\author{Carline Biesdorf}
\email{carline.fsc@gmail.com}
\affiliation{%
 Departamento de Física, CFM - Universidade Federal de Santa Catarina;  C.P. 476, CEP 88.040-900, Florianópolis, SC, Brasil 
}
\author{Debora P. Menezes}
\affiliation{%
 Departamento de Física, CFM - Universidade Federal de Santa Catarina;  C.P. 476, CEP 88.040-900, Florianópolis, SC, Brasil 
}%
\author{Luiz L. Lopes}
\affiliation{Centro Federal de Educa\c c\~ao  Tecnol\'ogica de
  Minas Gerais Campus VIII, CEP 37.022-560, Varginha, MG, Brasil}


\begin{abstract}

\textbf{Abstract}: In this paper the QCD phase diagram is 
obtained by the crossing of two types of effective models: the MIT bag based models are used to describe quark matter and QHD type models to describe hadronic matter, being the use of the former something new to this kind of approach and the latter used with improved parameterizations 
as compared with previous calculations. We use the Gibbs' conditions to establish the crossing points of the pressure in function of the chemical potential obtained in both phases.  We first analyze two-flavour symmetric matter constrained to both the freeze-out and the liquid-gas phase transition at the hadronic phase. Later we analyse the results for $\beta$-stable and charge neutral stellar matter and compare two different prescriptions: one that assumes flavour conservation, so that the quark phase is completely determined from the hadronic phase, a prescription never applied to finite temperatures before, and the other based on the Maxwell construction, where the quark phase is also $\beta$-stable. At the end, we compute the latent heat to find a signature of the critical end point.
\vspace{0.2cm}

  
\end{abstract}


\maketitle

\let\clearpage\relax

\section{Introduction}

It is currently understood that, at an elementary level, all matter is
made up of leptons and quarks. Quarks are subject to all four fundamental forces, as they have mass and electrical charge, which subject them to gravitational and electromagnetic force; they have a color charge, which subject them to the strong force, and can change flavour in decays, which is due to the weak force. Quantum chromodynamics (QCD) is the theory that explains the interaction between quarks through the strong force. 

The ordinary matter, however, is formed by hadrons. That is, under normal conditions, quarks are always confined inside hadrons. The reason behind confinement remains unknown and a one million dollar prize will be awarded for the correct explanation \cite{premio_confinamento}.
QCD has also a property called asymptotic freedom, 
proposed in 1973 by Frank Wilczek and David Gross \cite{gross1973ultraviolet}, and independently by David Politzer in the same year \cite{politzer1973reliable}, which says that the binding to which the quarks are subject decreases in intensity at small distances or high energies. This means that in extreme situations of high energy, such as the ones found in particle colliders or in the early universe, or at very high densities, such as the ones found inside compact stars, quarks can be deconfined. In fact, since the discovery of asymptotic freedom, the idea of the existence of a quark-gluon plasma (QGP) has been increasingly reinforced. The first experimental corroboration of the existence of the QGP occurred at RHIC in 2005 \cite{adams2005experimental}.

From the theoretical point of view, there are still many limitations for the QCD to be solved, which led physicists to attack the problem from two different perspectives: through QCD calculations on the lattice, the so called lattice QCD (LQCD), and through effective models. Due to computational constraints and numerical difficulties, such as the sign problem, the LQCD method only achieves results for low chemical potentials \cite{boricci2004reweighting,alexandru2005lattice,mendes2007lattice,bhattacharya2014qcd,goy2017sign}. This makes effective models currently the only source to generate results that cover the QCD phase diagram at higher chemical potentials.

The LQCD predicts the existence of a smooth crossover around a temperature of $160-170$~MeV at low chemical potentials, while the effective models predict a first order phase transition at higher densities \cite{aoki2006order,bellwied2015qcd,luostudy}. Moreover, the first order phase transition must end at an unique point where a second order phase transition takes place, the critical end point (CEP), even though its existence and exact location are not yet well established \cite{bazavov2017skewness,bazavov2017qcd}.

In a previous work \cite{lopes2021modified_pt2}, we have obtained an estimation of the QCD phase diagram based only on the extensions of the MIT bag model. Towards a more realistic description, in this work we use two relativistic effective models to describe the hadronic and quark phases of matter. 
For the quark phase, we use again an extended version of the MIT bag model, as introduced in ref.~\cite{lopes2021modified}, while for the hadronic phase, the quantum hadrodynamics (QHD)  model with non-linear terms and $\omega-\rho$ interaction is used, since it reconciles the theory with experimental results \cite{boguta/bodmer,PhysRevC.82.055803}. In both models the relative strength of the vector channel is fixed by symmetry group arguments, which allows us to build a unified scheme for the strong interaction. The Gibbs' conditions are used to establish the crossing points of the pressure as a function of the chemical potentials obtained in both phases \cite{landau2013course,glendenning2012compact}.  

To reproduce reliable results, some restrictions must be imposed when choosing the models mentioned above. The MIT based models are used only with constant values that don't allow stable u-d matter, and  QHD based models, in turn, are restricted to parameterizations that satisfy well known nuclear and astrophysical properties \cite{PhysRevC.90.055203,PhysRevC.93.025806}. Other model choices are certainly possible and depending (mainly) on the slope of the symmetry energy and other model properties, the phase transition points can be different. 

Similar works combining two effective models to construct the QCD phase diagram can be found in the literature, as for example \cite{di2011symmetry}, where the authors combined relativistic mean field and the MIT bag models to analyze symmetry energy effects on the mixed hadron-quark phase, and \cite{shao2011phase}, where the Nambu-Jona-Lasinio (NJL) model was used to describe the quark matter. NJL is a model that takes into account chiral symmetry aspects, which is not the case of the vector MIT bag model. In ref. \cite{lopes2021modified_pt2} we can see that they foresee different values for the phase transition line if it is computed with the quark matter model only. In the present study, the same can happen.

The modified vector MIT bag model used next to describe the quark matter and the QHD parametrizations consistent with recent astrophysical observations are new to this kind of approach. We also compare different prescriptions to obtain the crossing points, generally used in separate studies. Furthermore the use of these two models allows us to apply the same technique to both phases, once both models are relativistic and are dealt with at the mean field level. Although our results are model dependent, we do believe that the general trends would be the same had we chosen different models, despite the fact that the temperature and chemical potential values in all scenarios are not expected to be the same.

We first obtain phase diagrams for two-flavour symmetric matter, i.e., we take the hadronic matter to be constituted only by nucleons with equal chemical potentials $\mu_p=\mu_n$ and the quark matter constituted only by quarks $u$ and $d$ with $\mu_u=\mu_d$. We then confront these results with the chemical freeze-out line obtained by Cleymans \cite{cleymans2006comparison}, which is known to be a pure hadronic process, and, therefore, the freeze-out line needs to be in the hadronic phase. The liquid-gas phase transition \cite{finn1982nuclear}, which is also a pure hadronic process that takes place at low temperatures, must be confined to the hadronic part of the QCD phase diagram.

At $T=0$ we also expect the phase transition to occur at chemical potential values at least higher than $\mu=1050$~MeV, as shown in \cite{fukushima2010phase} using Polyakov loop formalism. In \cite{buballa2005njl,ruester2005phase}, using NJL-based models, it is shown that the chemical potential of the phase transition at $T=0$ occurs in the range $\mu=1080 - 1100$~MeV. In \cite{klahn2017simultaneous}, using a chiral bag model, the obtained value is $\mu=1250$~MeV. Although there is no experimental evidence of the maximum chemical potential which preserves the hadron phase, in \cite{annala2020evidence} it is pointed out that quark matter inside neutron stars is not only possible, but probable. For $\beta$-stable matter, using a relativistic density functional approach, \cite{ayriyan2018robustness} indicates that the transitions should occur around $\mu=1200$~MeV. As done in our previous work \cite{lopes2021modified_pt2}, for two-flavour symmetric matter we assume a maximum value of $\mu=1400$~MeV as a more conservative estimate.

Next we obtain the phase diagram for stellar matter, i.e. $\beta$-stable and charge neutral matter, and compare two different prescriptions that affect the quark phase, one where we impose flavour conservation during the phase transition, so that the quark matter is totally determined from the hadronic matter \cite{bombaci2017quark}
and the other where the quark matter is also $\beta$-stable and charge neutral, the so called Maxwell prescription, usually taken to build hybrid stars \cite{voskresensky2002charge,paoli2010importance}. The prescription of flavour conservation has already been used for T=0 \cite{bombaci2017quark,JCAP12(2017)028_Kauan}, but for finite temperatures relevant to protoneutron star evolution, it is something completely new.


It is important to stress that the prescriptions we have just mentioned can only provide first order phase transitions since they depend on relativistic models within mean field approximations (MFA) and quantum fluctuations are completely disregarded. 
Another aspect is that one should be aware of the limitations imposed by extrapolations of stellar matter conditions to high temperatures. For completeness we analyse the dependence of the results on the models used at both ends of the curve, i.e, high temperatures and low chemical potentials and low temperatures and high chemical potentials, but our results at low chemical potentials and high temperatures serve as a guide only and should be regarded with a grain of salt. 

Finally, we calculate the latent heat based on two different simple expressions to try to estimate the position (temperature and corresponding chemical potential) of the critical end point.

\section{Effective models}\label{sec:the_effective_models}

\subsection{Hadronic matter - Relativistic Mean Field QHD models}

To describe hadronic matter we use the Walecka Model \cite{NLWM} with non-linear terms \cite{boguta/bodmer}. 
In this model four types of mesons are included to describe the interactions between baryons; so we have the scalar $\sigma$, isoscalar-vector $\omega^\mu$, isovector-vector $\vec{\rho}^\mu$  and strange isoscalar-vector $\phi^\mu$ meson fields. As in \cite{PhysRevC.84.065810,PhysRevLett.95.122501,PhysRevC.82.025805,PhysRevC.82.055803} we also consider the $\omega$-$\rho$ meson coupling terms as this term influences the symmetry energy and its slope, resulting in Equations of State (EoSs) that can satisfy all important nuclear matter saturation properties and observational constraints. The inclusion of the $\phi$ meson does not affect the properties of the nuclear matter as it does not couple to the nucleons, but it stiffens the EoSs that include hyperons.  The Lagrangian density is as follows:

\begin{widetext}
\begin{center}

\begin{align}\label{NLWM}
    \mathcal{L}_{NLWM}& = \sum_{B}\overline{\psi}_B [\gamma_\mu(i\partial^\mu - g_{B \omega}\omega^\mu - g_{B \rho} \frac{\vec{\tau}_B}{2} \vec{\rho}^\mu - g_{B \phi} \phi^\mu)
    -m^*_B]\psi_B
    +\frac{1}{2}\partial_\mu \sigma \partial^\mu \sigma - \frac{1}{2} m_\sigma^2 \sigma^2 - \frac{1}{3!} \kappa \sigma^3 - \frac{1}{4!} \lambda \sigma^4 + \nonumber \\
    &+\frac{1}{2} m_\phi^2 \phi_\mu \phi^\mu -\frac{1}{4} \Omega^{\mu\nu}\Omega_{\mu\nu} + \frac{1}{2}m_\omega^2 \omega_\mu \omega^\mu 
    - \frac{1}{4}\Phi^{\mu \nu} \Phi_{\mu \nu} 
     - \frac{1}{4} \vec{R}_{\mu \nu} \vec{R}^{\mu \nu} + \frac{1}{2}m_\rho^2 \vec{\rho}_\mu \vec{\rho}^\mu + \Lambda_v g_{N \omega}^2 g_{N \rho}^2\omega_\mu \omega^\mu \vec{\rho}_\mu \vec{\rho}^\mu, 
\end{align}
\end{center}
\end{widetext}
where the Dirac spinor $\psi_B$ represents the baryons with the effective mass $m_B^*=m_B-g_{B \sigma} \sigma$, $\vec{\tau}_B$ are the corresponding Pauli matrices, $g_{Bi}$ are the coupling constants of the mesons $i=\sigma, \omega, \rho, \phi$ with the baryon $B$, $m_i$ is the mass of the meson $i$, $\Omega_{\mu \nu}=\partial_\mu \omega_\nu - \partial_\nu \omega_\mu$, $\vec{R}_{\mu \nu}=\partial_\mu\vec{\rho}_\nu - \partial_\nu\vec{\rho}_\mu - g_\rho(\vec{\rho}_\mu \times \vec{\rho}_\nu)$ and $\Phi_{\mu \nu}=\partial_\mu \phi_\nu - \partial_\nu \phi_\mu$. $\kappa$ and $\lambda$ are scalar self-interaction constants introduced by \cite{boguta/bodmer}
and $\Lambda_v$ is the coupling constant of the mixed quartic isovector-vector interaction. The $B$ sum extends over the octet of the lightest baryons $\{n, p, \Lambda, \Sigma^-, \Sigma^0, \Sigma^+, \Xi^-, \Xi^0\}$ or only over the nucleons, depending on the scenario considered (see section \ref{sec:Conditions}).


After applying the mean field approximation (MFA), the EoS can be easily obtained from Eq.~(\ref{NLWM}) and detailed calculations can be seen in \cite{PhysRevC.90.055203}, for instance. 

As it is well known, it is easy to find in the literature several parametrizations 
for this model, but not all of them satisfy the constraints imposed by experimental results for nuclear matter, as shown in \cite{PhysRevC.90.055203}. Moreover, it is desirable that the EoSs reproduce neutron star masses that are heavier than $2$~M$_\odot$, as imposed by the NICER results for PSR J$0740 + 6620$ ($2.08 \pm 0.07$M$_\odot$ M$_\odot$) \cite{miller2021radius} and the recently detected PSR J0952-0607, with a mass of $2.35 \pm 0.17$M$_\odot$\cite{romani2022psr}, yet to be confirmed.

In the present work we choose two parametrizations: eL3$\omega \rho$ and NL3$^*\omega \rho$. The eL3$\omega \rho$ is the parametrization proposed in \cite{lopes2021hyperonic} with a slightly modification to adjust the symmetry energy and its slope accordingly to \cite{essick2021astrophysical} and the NL3$^*\omega \rho$ is the NL3$^*$ parametrization proposed in \cite{lalazissis2009effective} with the addition of the $\omega \rho$-channel as done in \cite{lopes2022nature}. All of them satisfy the symmetric nuclear matter properties at the saturation density and also reproduce maximum star masses above $2$~M$_\odot$ even when hyperons are included. The main parameters of these parametrizations are presented in Table~\ref{tab:parametros} and the main nuclear properties are presented in Table~\ref{tab:propriedades}.

\begin{widetext}
\begin{center}
\begin{table}[]
\caption{
Parameter sets for the two models discussed in the text. The meson masses $m_\sigma$, $m_\omega$, and $m\rho$ are all given in MeV. The nucleon and the $\phi$ meson masses were fixed at $M=939$ MeV and $m_\phi=1020$~MeV, respectively, in both models.}
\begin{tabular}{p{1.5cm}p{1.2cm}p{1.2cm}p{1.2cm}p{1.5cm}p{1.5cm}p{1.5cm}p{1.9cm}p{2.2cm}p{1.2cm}}
\toprule[0.6pt]
 Model  & $m_\sigma$  & $m_\omega$  & $m_\rho$  & $g_{N \sigma}$  & $g_{N \omega}$  & $g_{N \rho}$  & $\kappa$  & $\lambda$  & $\Lambda_v$  \\ \midrule[1.5pt]
 eL3$\omega \rho$ & $512.000$  & $783.000$  & $770.000$  & $9.0286$  & $10.5970$  & $9.4381$  & $0.008280 \cdot g_\sigma^3$  & $-0.023400 \cdot g_\sigma^4$  & $0.0283$  \\
 NL3$^*\omega \rho$ & $502.574$  & $782.600$  & $763.000$  & $10.0944$  & $12.8065$  & $14.4410$  & $0.004417\cdot g_\sigma^3$  & $-0.017422\cdot g_\sigma^4$  & $0.045$ \\ \bottomrule[0.6pt]
\end{tabular}\label{tab:parametros}
\end{table}
\end{center}
\end{widetext}

\begin{table}[]
\caption{Properties at saturation of the models eL3$\omega \rho$ and NL3$^*\omega \rho$. We present the saturation density (n$_0$), energy per particle (E/A), compressibility (K), and effective nucleon mass (M$^*$/M) in symmetric nuclear matter, and also the symmetry energy (E$_{sym}$) and slope of the symmetry energy (L) at n$_0$.}
\begin{tabular}{p{2.0cm}p{1.4cm}p{1.4cm}}
\toprule[0.6pt]
 Model  & eL3$\omega \rho$  & NL3$^*\omega \rho$    \\ \midrule[1.5pt]
 n$_0$~(fm$^{-3}$)  & 0.156  & 0.150   \\ 
 E/A (MeV)    & 16.2   &    16.3       \\
 K (MeV)  & 256 & 258     \\
 M$^*$/M & 0.69  & 0.59    \\
 E$_{sym}$~(MeV) & 32.1  & 30.7   \\ 
 L (MeV)    &   66   &  42     \\\bottomrule[0.6pt]
\end{tabular}\label{tab:propriedades}
\end{table}

We consider the hyperon masses to be $m_\Lambda=1116$~MeV, $m_{\Sigma}=1193$~MeV and $m_{\Xi}=1318$~MeV. The couplings of the hyperons to the vector mesons are related to the nucleon couplings, $g_{N \omega}$ and $g_{N \rho}$, by assuming SU(6)-flavour symmetry, according to the ratios \cite{dover1984hyperon,schaffner1996hyperon,banik2014new,tolos2017equation,lopes2014hypernuclear}:

\begin{align}
    &g_{\Lambda \omega} : g_{\Sigma \omega} : g_{\Xi \omega} : g_{N \omega} = \frac{2}{3} : \frac{2}{3} : \frac{1}{3} : 1, \nonumber \\
    &g_{\Lambda \rho} : g_{\Sigma \rho} : g_{\Xi \rho} : g_{N \rho} = 0:2:1:1,\\
    &g_{\Lambda \phi} : g_{\Sigma \phi} : g_{\Xi \phi} : g_{N \omega} = -\frac{\sqrt{2}}{3} : -\frac{\sqrt{2}}{3} : -\frac{2\sqrt{2}}{3} : 1 \nonumber,
\end{align}
noting that $g_{N \phi}=0$. The coupling of each hyperon to the $\sigma$ field is adjusted to reproduce the hyperon potential in strange nuclear matter (SNM) derived from hypernuclear observables. We fix this potentials as $U_\Lambda(n_0)=-28$~MeV, $U_\Sigma(n_0)=+30$~MeV and $U_\Xi(n_0)=-4$~MeV and obtain the coupling constants presented in Table~\ref{tab:acoplamentos}.

\begin{table}[]
\caption{Hyperon-$\sigma$ coupling constants adjusted to reproduce the hyperon potential in SNM derived from hypernuclear observables.}
\begin{tabular}{p{1.5cm}p{1.4cm}p{1.4cm}p{1.4cm}}
\toprule[0.6pt]
 Model  & $g_{\Lambda \sigma}/g_{N \sigma}$  & $g_{\Sigma \sigma}/g_{N \sigma}$  & $g_{\Xi \sigma}/g_{N \sigma}$  \\ \midrule[1.5pt]
 eL3$\omega \rho$  & $0.610$  & $0.406$  & $0.269$ \\ 
 NL3$^*\omega \rho$ & $0.613$  & $0.461$  & $0.279$  \\ \bottomrule[0.6pt]
\end{tabular}\label{tab:acoplamentos}
\end{table}


\subsection{Quark matter - MIT bag based models}

To describe the quark matter we choose the simple MIT bag model \cite{PhysRevD.9.3471_mit_original} and a modification of that model that includes a vector field as presented in \cite{lopes2021modified}. In the original MIT bag model the quarks are free inside the bag and confined inside it. All the information about the strong force relies on the bag constant $B$, which mimics the vacuum pressure. With the inclusion of a vector field $V_\mu$, the quark interaction inside the bag is mediated by the $\omega$ meson, in a similar way as in the QHD model with the baryons. In \cite{lopes2021modified} a self-interacting vector field which allows more malleability on the stiffness of the EoSs has also been added. However, in the present study, we opted not to include this term as its influence is barely noticeable for relatively low densities. 
The Lagrangian density of the model follows:

\begin{align}\label{mit}
    \mathcal{L}_{MIT}&=\sum_q \Big\{ \overline{\psi}_q \Big[\gamma^\mu(i \partial_\mu -g_{qqV} V_\mu) - m_q\Big]\psi_q   \nonumber \\
    &+ \frac{1}{2}m_V^2 V_\mu V^\mu - B\Big\}\Theta(\overline{\psi}_q \psi_q) - \frac{1}{2}\overline{\psi}_q \psi_q \delta_S,
\end{align}
where the Dirac spinor $\psi_q$ represents the quark with mass $m_q$, $g_{qqV}$ the coupling constant, and $m_V$ the mass of the meson. $\Theta(\overline{\psi}_q \psi_q)$ is a Heaviside function that ensures that the quarks are confined inside the bag and $\delta_S$ is a Dirac function that guarantees continuity of the fields of the quarks on the surface of the bag. If we take $g_{qqV}=m_V=0$ we obtain the original MIT bag model. Using mean field approximation we easily obtain the EoS. For more details, the interested reader can see reference \cite{lopes2021modified}.

In all cases we choose the quark masses as being $m_u=m_d=4$~MeV and $m_s=95$~MeV \cite{tanabashi2018review}. The meson mass $m_V$ 
is the same as the $\omega$ meson mass $m_\omega$ considered in the QHD models.

As we have done in the hadronic phase, for the relation between the coupling constants we opt to use the ones obtained via symmetry group, which allows us to build an unified scheme for the strong interaction. In such approach we have 
$$g_{ssV}=\frac{2}{5} g_{uuV}=\frac{2}{5} g_{ddV},$$
as done in \cite{lopes2021modified}. We also define $G_V=(g_{uuV}/m_V)^2$ and choose here $G_V=0.05$~fm$^2$, $G_V=0.1$~fm$^2$ and $G_V=0.3$~fm$^2$.

As for the bag pressure value, we choose a temperature-dependent bag model in order to be able to obtain higher transition temperatures at low chemical potentials and at the same time maintain the transition chemical potentials at low temperatures within a certain range. More details on this discussion can be seen in \cite{lopes2021modified_pt2} and \cite{biesdorf2022qcd}. So, the $B$ in Eq. \ref{mit} is substituted by:

\begin{equation}\label{eq:B(T)}
    B(T)=B_0 \Bigg[1+\left(\frac{T}{T_0} \right)^4 \Bigg],
\end{equation}
where $T_0$ is adjusted to reproduce the LQCD and freeze-out (pseudo) critical temperature at zero chemical potential. Thus we use $T_0=131$~MeV for $B_0^{1/4}=148$~MeV and $T_0=155$~MeV for $B_0^{1/4}=165$~MeV. As for the values of $B_0$, we choose $B_0^{1/4}=148$~MeV, which is the lowest value within the stability window \cite{torres2013quark,lopes2021modified} that satisfies the Bodmer-Witten conjecture \cite{bodmer1971collapsed,witten1984cosmic}, and $B_0^{1/4}=165$~MeV, which is outside the stability window, but allows us to satisfy another constraint that will become clear latter on.

\section{Conditions for phase coexistence - Gibbs' conditions}

Two states of matter which can exist simultaneously in equilibrium with each other and in contact are described as different phases. The equilibrium conditions can be used to determine when the transition from one phase to another occurs. For the two phases to be in equilibrium it is necessary that, first of all, their temperatures are the same. Also, as the forces exerted by the two phases on each other at their surface of contact must be equal and opposite, the pressures have also to be equal. And, finally, the chemical potentials have to be identical \cite{landau2013course}. 

As in our case the two phases considered are the hadronic phase ($H$) and the quark phase ($Q$), we can write:

\begin{align}\label{condicoes_transicao_HQ}
	& T^{H} = T^{Q} = T,\nonumber\\
	& P^{H} = P^{Q} = P_0 \\
	& \mu^{H}(P_0,T) = \mu^{Q}(P_0,T) = \mu_0,\nonumber
\end{align}


These are the necessary conditions for thermodynamic equilibrium of the hadronic and quark phases, also called Gibbs' conditions.  The detailed calculation of the relevant physical quantities can be easily  found in the literature for both the hadronic (ref.~\cite{NLWM,PhysRevC.90.055203,shao2011phase}) and the quark (ref.~\cite{lopes2021modified,lopes2021modified_pt2} phases. The total baryonic chemical potential for the hadron ($\mu^H$) and the quark ($\mu^Q$) phases are given in terms of its individual constituents~\cite{lopes2022hypermassive}:

\begin{eqnarray}
\mu^H_B = \frac{(\sum_H n_H \mu_H + \sum_l n_l \mu_l)}{\sum_H n_H}, \nonumber \\
\mu^Q_B = \frac{3(\sum_Q n_Q \mu_Q + \sum_l n_l \mu_l)}{\sum_Q n_Q},  \label{chempot}
\end{eqnarray}
where $H$, $l$ and $Q$ stand for hadrons, leptons and quarks, respectively.
It is also worth emphasizing that due to the model limitations and the prescriptions we use, Eq.~\ref{condicoes_transicao_HQ} gives rise to
a first order phase transition, while a more realistic approach should reproduce a smooth crossover at low chemical potentials~\cite{aoki2006order,bellwied2015qcd}. 
In the last section of this paper, we analyse possible ways to limit the temperature and chemical potentials at which the critical end point is reached and hence, define the limits of our calculation.

\begin{figure}[ht] 
\begin{centering}
 \includegraphics[angle=0,width=0.45\textwidth]{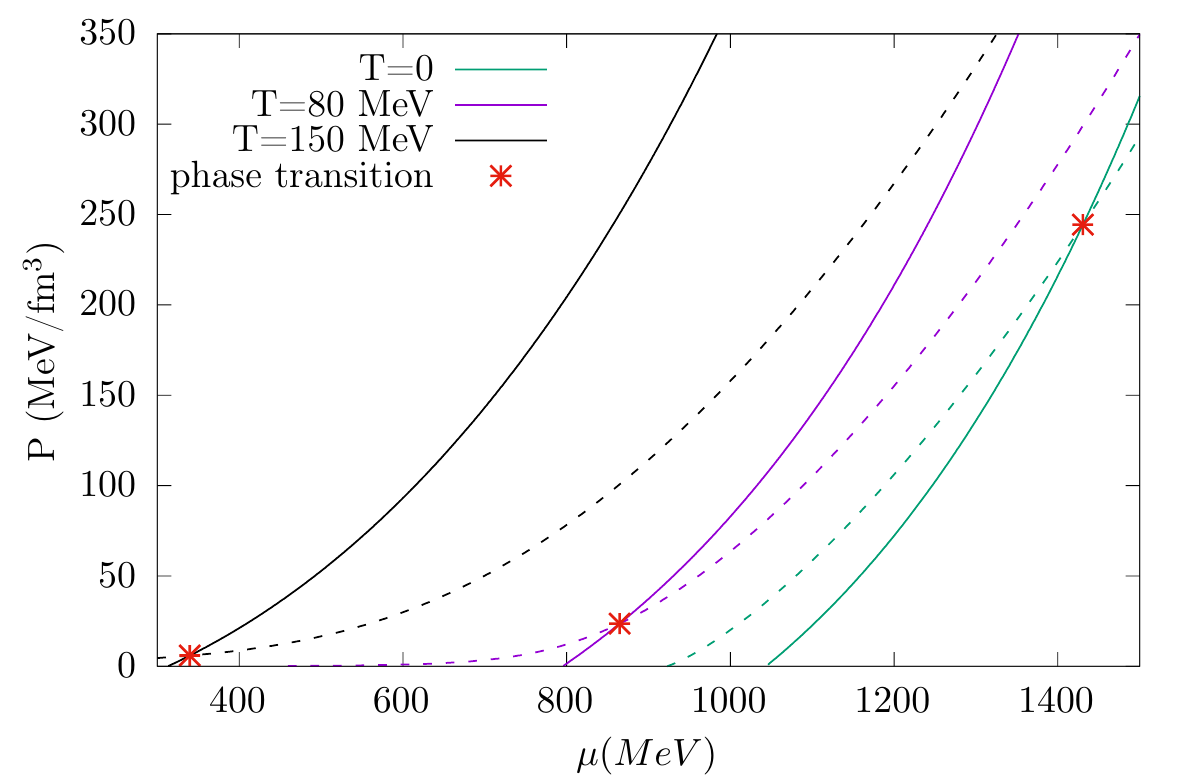}
\caption{Relation between pressure and chemical potential for the hadron (dashed lines) and quark (solid lines) phases, respectively, described by the eL3$\omega \rho$ parametrization and $B_0^{1/4}=165$~MeV considering three different temperatures. Both phases are of symmetric matter. The red dots are the points where the Gibbs' conditions are satisfied.}  \label{fig:cruzamento}
\end{centering}
\end{figure}

In Fig.~\ref{fig:cruzamento} we show examples where the Gibbs' conditions are met for various temperatures. In this case we use symmetric matter for both phases, with $\mu_n=\mu_p$ for the hadron phase and $\mu_u=\mu_d$ for the quark phase, but in the present work we will also explore constructions using stellar matter, i.e., a charge neutral matter in $\beta$-equilibrium, as already explained in the Introduction of the paper. In all cases, the phase transition points are found as shown in Fig.~\ref{fig:cruzamento}.

\section{Different matter hypotheses}\label{sec:Conditions}
Here we discuss the conditions that we impose to the hadronic matter before it goes through a first order phase transition and also the sort of conservation that we impose when the transition occurs. 

\subsection{Based on symmetric matter}\label{subsec:Symmetric_matter}

The first situation  we consider is a two-flavour symmetric matter where the nucleons have equal chemical potentials ($\mu_n=\mu_p$) as well as the quarks ($\mu_u=\mu_d$). 

The main reason to consider this scenario is to adjust our parametrizations in order to satisfy the constraints imposed by the chemical freeze-out line \cite{cleymans2006comparison}, which, as being a pure hadronic process, needs to be in the hadronic phase.

\subsection{Based on stellar matter}\label{subsec:Stellar_matter}

Stable neutron and quark stars are normally described by zero temperature matter, but temperatures up to $T=50$ MeV can be important in the early moments of the star as the cooling of a newborn neutron star by neutrino diffusion takes a few seconds \cite{menezes2004warm}. For higher temperatures we do not expect $\beta$-stable matter \cite{gupta2003study,cavagnoli2008warm}, nevertheless here we extend our study beyond the temperature of $T=50$ MeV for educational purposes.

For a description of neutral and stable stellar matter we include a non-interacting lepton gas to both the hadronic matter and the quark matter. For that, we add the following Lagrangian density to Eqs.~(\ref{NLWM}) and (\ref{mit}):

\begin{equation}\label{leptons}
    \mathcal{L}_l=\sum_l \overline{\psi}_l (i \gamma^\mu \partial_\mu - m_l)\psi_l,
\end{equation}
where the sum extends over the leptons $e^-$ and $\mu^-$ with mass $m_{e^-}=0.511$ MeV and $m_{\mu^-}=105.66$ MeV, respectively \cite{tanabashi2018review}.

Furthermore, as neutron stars present internal densities that can be up to 10 times higher that the nuclear saturation density, the onset of hyperons is expected because their appearance is energetically favorable as compared with the inclusion of more nucleons in the system \cite{menezes2021neutron}, and so, here we add the six lightest hyperons to hadron matter so that the sum in Eq.~(\ref{NLWM}) extends over the baryon octet. Nevertheless, at the end of this work we also obtain diagrams for stellar matter without strangeness so then, in this case, the sum in Eq.~(\ref{NLWM}) extends only over the nucleons.

We also have to impose $\beta$-equilibrium and electric charge neutrality to the hadronic matter:

\begin{equation}
    \mu_B=\mu_n-q_B ~\mu_e \quad \textrm{and} \quad \mu_{e^-}=\mu_{\mu^-},
\end{equation}

\begin{equation}
    n_p + n_{\Sigma^+} = n_{e^-} + n_{\mu^-} + n_{\Sigma^-} + n_{\Xi^-}.
\end{equation}

Next we consider two scenarios. In the first one we impose flavour conservation at the point of the phase transition, so that the quark phase is completely determined from the initial hadronic matter through the bond:

\begin{equation}\label{eq:charge_conser}
    y_q=\frac{1}{3}\sum_i n_{qi} y_i,
\end{equation}
where $i=n,p,\Lambda,\Sigma^-,\Sigma^0,\Sigma^+,\Xi^-,\Xi^0$, $y_i=n_i/n_B$, being $n_i$ the baryon density of baryon $i$ and $n_B$ the total baryon density, and $n_{qi}$ the number of quarks with flavour $q$ that constitute baryon $i$ \cite{olesen1994nucleation}. 
Thus, it is generally assumed that under certain circumstances, the electrically neutral and in chemical equilibrium hadronic matter is metastable and can be converted into an energetically favored deconfined quark phase. Due to the imposition shown in Eq.~\ref{eq:charge_conser}, this matter will not be in $\beta$-equilibrium \cite{JCAP12(2017)028_Kauan,bombaci2017quark}.

Within this scenario we also briefly analyze two sub-scenarios. One where the lepton fraction, defined as the fraction between lepton density and the baryonic density of each phase is preserved at the point of the phase transition and the other where we have the same lepton density, and consequently same energy density and pressure, for both phases at the point of the hadron-quark phase transition.


We also consider a second scenario, where we do not impose flavour conservation, but charge neutrality and chemical equilibrium to both phases, so that we have, for the quark matter:

\begin{equation}
    \mu_s=\mu_d=\mu_u+\mu_e \quad \textrm{and} \quad \mu_{e^-}=\mu_{\mu^-},
\end{equation}

\begin{equation}
    n_e+n_\mu=\frac{1}{3}(2n_u-n_d-n_s).
\end{equation}

This second scenario is the so called Maxwell construction (or prescription), most commonly used to construct the EoSs to describe hybrid stars, as done, for instance, in~\cite{lopes2022hypermassive}.

\begin{figure*}[ht] 
\begin{centering}
 \includegraphics[angle=0,width=1.0\textwidth]{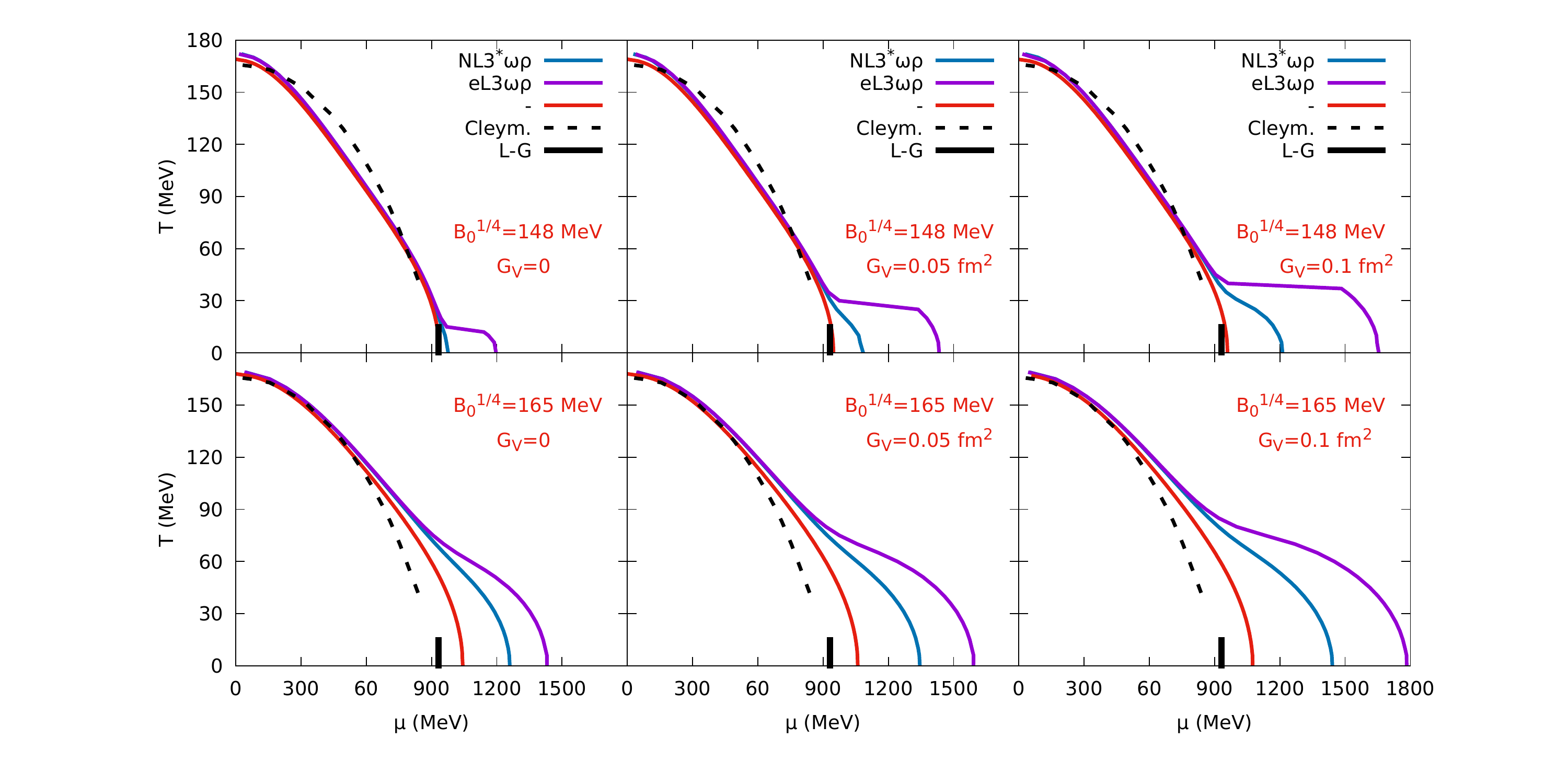}
\caption{Phase diagrams for two-flavour symmetric matter considering the NL3$^*\omega \rho$ and eL3$\omega \rho$ parametrizations for the hadronic matter and different MIT bag based models for two temperature dependent bag $B(T)$ values for the quark matter. The Cleym. line is the experimental freeze-out \cite{cleymans2006comparison} and L-G is the region where we expect a liquid-gas phase transition \cite{finn1982nuclear}. The red lines correspond to the results obtained with the MIT based bag models only, where the phase transition criterion is just the value of the chemical potential where the pressure goes to zero, as done in \cite{lopes2021modified_pt2}.}
 \label{fig:symmetric_matter}
\end{centering}
\end{figure*}

\section{Phase Diagrams}

In this section we present the results first for two-flavour symmetric matter and then for stellar matter.

\subsection{Symmetric Matter}\label{sec:results-Sym_matter}

In Fig.~\ref{fig:symmetric_matter} we present the phase diagrams obtained considering two-flavour symmetric matter for both phases, with $\mu_n=\mu_p$ and $\mu_d=\mu_u$. In each graphic we present three curves, one where we use the NL3$^* \omega \rho$ parametrization (blue line) and other where we use the eL3$\omega \rho$ parametrization (purple line) in combination with the MIT bag model and we also present the curve obtained using only the MIT bag model (red line), i.e., at this last one, instead of constructing the phase diagram with two models, quark matter pressure is simply set to be zero (see ref. \cite{lopes2021modified_pt2}, for details). The graphics at the top are obtained using a bag pressure value $B_0^{1/4}=148$~MeV and for the ones at the bottom we use $B_0^{1/4}=165$~MeV. From left to right we vary the value of the vector channel of the MIT bag model where we use $G_V=0.0$, $G_V=0.05$~fm$^2$ and $G_V=0.1$~fm$^2$.

We first take a look at the region of low chemical potentials. As can be seen, all curves converge more or less to the same point, so that the maximum temperatures (by maximum temperature $T_{max}$ we mean merely the highest temperature where we are able to satisfy the phase transition conditions at $\mu \rightarrow 0$) obtained are above the freeze-out line. This happens because we choose the values of $T_0$ exactly to satisfy the restrain imposed by the LQCD and freeze-out results, which states that the temperatures at very low chemical potentials should be around  $T=168$~MeV. In \cite{biesdorf2022qcd} we have already verified that the maximum temperature obtained depends only on the bag pressure value: the higher the $B^{1/4}$, the higher the maximum temperature. The same conclusion can be drawn here. 
Following, it is interesting to note that the low chemical potential
results obtained using the combination of the two models are the same as the results using only the quark model. That is because the Gibbs' conditions at this region are met at very low pressures ($P_0 \leq 5$~MeV/fm$^3$), as can be seen in Fig. \ref{fig:cruzamento}, and so, as the condition for phase transition when using only the MIT bag model is $P=0$, it comes with no surprise that this results are very much alike. The same is true when we use a bag pressure value with no dependence on the temperature, as done in \cite{biesdorf2022qcd}.

Now we look at the region of the diagrams of high chemical potentials, i.e., $T \rightarrow 0$. We see that the results are different for each parametrization. Let us first analyze the effect of the different MIT bag based model parametrizations. The most evident effect is that a higher bag pressure value results in a higher chemical potential. This follows what happens when using only the MIT model, a case easy to understand: in the equation for the pressure, the value of the Bag subtracts from the value of the pressure of the quarks (likewise, the meson $\omega$ when $G_V \neq 0$) and so, the higher the Bag value the higher must the contribution for the pressure coming from the quarks be in order to meet the condition for phase transition ($P=0$) and, consequently, the chemical potential is also higher. When we use two models the increase in the chemical potential is even greater because the Gibbs' conditions are met at pressures higher than zero ($P_0 > 0$) so that the pressure realized by the quarks and $\omega$ meson is even higher, which also increases the chemical potential at the transition point $\mu_0$.

The inclusion and increase of the $G_V$ value also follows what happens when we use only one model, the higher its value, the higher the chemical potential. As already stated in \cite{lopes2021modified_pt2}, the vector field causes an additional repulsion between the quarks, increasing the chemical potential of the phase transition. And again, as $P_0 > 0$, the effect when using two models is increased, so much so that here we use smaller values for $G_V$ than used in our previous work \cite{lopes2021modified_pt2}.

As for the influence of the different parametrizations of the QHD based model, we can see that the eL3$\omega \rho$ parametrization always results in higher chemical potentials, and, as the NL3$^* \omega \rho$ is a stiffer EoS, we conclude the same as already noticed in \cite{biesdorf2022qcd}, a stiffer hadron EoS results in a lower chemical potential. 

The behavior of the diagrams at low temperatures is not so smooth, as we can observe an abrupt increase of the chemical potential, especially when using a bag pressure value of 148 MeV. This happens because the lower the temperature, the smaller the change in the graphs of $\mu$ x P for different temperatures. And more so, at lower temperatures the curves for the hadron and the quark phases are very close to each other. As a result, a small change in the temperature makes the crossing point of the lines take a big leap resulting in an abrupt change in the chemical potential on the phase diagram.


In Table \ref{tab:pot_max} we present the values of the maximum chemical potential obtained for each combination of parametrizations used in Fig.~\ref{fig:symmetric_matter}. By maximum chemical potential $\mu_{max}$ we mean the chemical potential where a phase transition at $T=0$ takes place.
Only a few combinations of parametrizations result in a chemical potential within the range $1050 \leq \mu \leq 1400$~MeV, namely: NL3$^*\omega \rho$ in combination with $B^{1/4}=148$~MeV + $G_V=0.05$~fm$^2$ and $G_V=0.1$~fm$^2$ and $B^{1/4}=165$~MeV + $G_V=0.0$ and $G_V=0.05$~fm$^2$ and eL3$\omega \rho$ in combination with $B^{1/4}=148$~MeV. When using only the MIT based model the results for $B^{1/4}=165$~MeV + $G_V=0.05$~fm$^2$ and $G_V=0.1$~fm$^2$ are also within this range.

\begin{widetext}
\begin{center}
\begin{table}[]
\caption{Chemical potentials ($\mu_{max}$) at $T=0$ for all combinations of parametrizations considering \textbf{symmetric} matter. The first line corresponds to the results obtained with the MIT based bag models only, where the phase transition criterion is just the value of the chemical potential where the pressure goes to zero, as done in \cite{lopes2021modified_pt2}.
All values are given in MeV }
\begin{tabular}{p{1.5cm}|p{2.5cm}|p{2.5cm}|p{2.5cm}|p{2.5cm}|p{2.5cm}|p{2.5cm}}
\toprule[0.6pt]
 Models  & $B_0^{1/4}=148$~MeV  & $B_0^{1/4}=148$~MeV + $G_V=0.05$~fm$^2$  & $B_0^{1/4}=148$~MeV + $G_V=0.1$~fm$^2$ & $B_0^{1/4}=165$~MeV  &   $B_0^{1/4}=165$~MeV + $G_V=0.05$~fm$^2$    &   $B_0^{1/4}=165$~MeV + $G_V=0.1$~fm$^2$   \\ \midrule[1.5pt]
 - & $\mu_{max}=936$  & $\mu_{max}=948$  & $\mu_{max}=959$  & $\mu_{max}=1043$ &   $\mu_{max}=1060$ &   $\mu_{max}=1075$  \\
 NL3$^*\omega \rho$ & $\mu_{max}=976$  & $\mu_{max}=1084$  & $\mu_{max}=1212$  & $\mu_{max}=1260$   &   $\mu_{max}=1345$    &   $\mu_{max}=1441$  \\
 eL3$\omega \rho$  & $\mu_{max}=1197$  & $\mu_{max}=1434$  & $\mu_{max}=1655$  & $\mu_{max}=1431$    &   $\mu_{max}=1592$    &   $\mu_{max}=1783$ \\\bottomrule[0.6pt]
\end{tabular}\label{tab:pot_max}
\end{table}
\end{center}
\end{widetext}

As we chose $T_0$ in order to obtain higher temperatures at low chemical potentials, we are able to obtain a maximum temperature around $T=168$~MeV in all combinations of parametrizations, but we are only able to fit the freeze-out line entirely inside the confined (hadron) phase for the combinations that include a $B_0^{1/4}=165$~MeV and a QHD based model. Which means that we were able to obtain only two combinations of parametrizations that satisfy both constraints for low and high temperatures: NL3$^*\omega \rho$ + $B_0^{1/4}=165$~MeV and NL3$^*\omega \rho$ + $B_0^{1/4}=165$~MeV + $G_V=0.05$~fm$^2$.


\subsection{Stellar Matter}\label{sec:results-Stel_matter}

\begin{figure*}[ht] 
\begin{centering}
 \includegraphics[angle=0,width=1.0\textwidth]{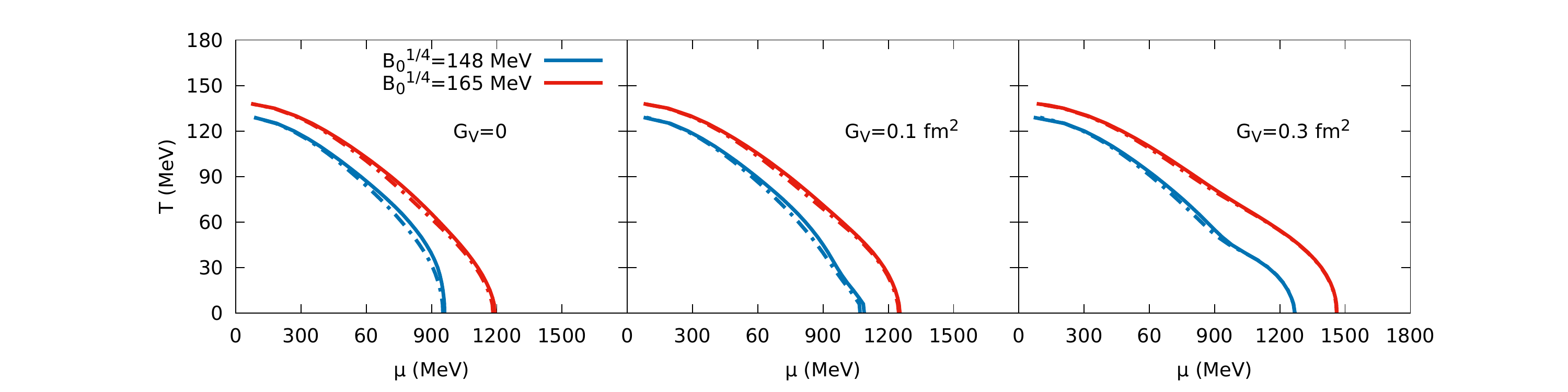}
\caption{Comparison between two different prescriptions for the lepton matter in the quark phase for strange stellar matter considering flavour conservation. Here we present the results for the NL3$^*\omega \rho$ and different parametrizations for the MIT based model. The solid lines represent the prescription of lepton fraction conservation and the dashed lines represent the prescription of lepton density conservation. }\label{fig:lepton_conser}
\end{centering}
\end{figure*}

Here we present the results for charge neutral matter in $\beta$-equilibrium for two scenarios, one considering flavour conservation, which results in quark matter that is not in $\beta$-equilibrium, and another with no flavour conservation where both phases are in $\beta$-equilibrium, a typical Maxwell prescription used to construct hybrid stars, as done in \cite{lopes2021hyperonic}, for example. But first we analyze the results for a sub-scenario within the scenario of flavour conservation, where we analyze different prescriptions for the lepton matter during the phase transition. In the first one, the lepton fraction, defined as the fraction between lepton density and the baryonic density of each phase is preserved at the point of the phase transition and in the other one, the lepton density is taken to be the same, and consequently the same energy density and pressure are obtained, for both phases at the point of the hadron-quark phase transition. In Fig.~\ref{fig:lepton_conser} we compare the two prescriptions. 


As the different prescriptions only affect the quark phase, we choose here to present the results from only one of the QHD based model, namely, the NL3$^* \omega \rho$, and vary the MIT based model parametrizations. As can be seen, the differences are very small. They are greater for the bag pressure value of $B_0^{1/4}=148$~MeV (blue lines) than for the $B_0^{1/4}=165$~MeV (red lines), and also decrease with the increase of the $G_V$ constant. But for all parametrizations the prescription considering the lepton fraction equal at both phases at the transition point (solid lines) is the one where the chemical potentials are always slightly higher. Because of the imposition of flavour conservation and the Gibbs' conditions as criteria for the phase transition, there is an increase of the baryon density at the point of the phase transition, so, in the sub-scenario where we impose the lepton fraction conservation this means also an increase, in the same proportion, of the lepton density, which, in turn, means an increase in the chemical potential. As the contribution to the chemical potential from the leptons is very small, these differences are also very small.

From this sub-scenario we choose the one where there is a conservation of the lepton fraction at the point of the phase transition because, in this case, the charge neutrality is also preserved. This approach is contrary to what was done in \cite{pelicer2022phase}. 

\begin{figure*}[ht] 
\begin{centering}
 \includegraphics[angle=0,width=1.0\textwidth]{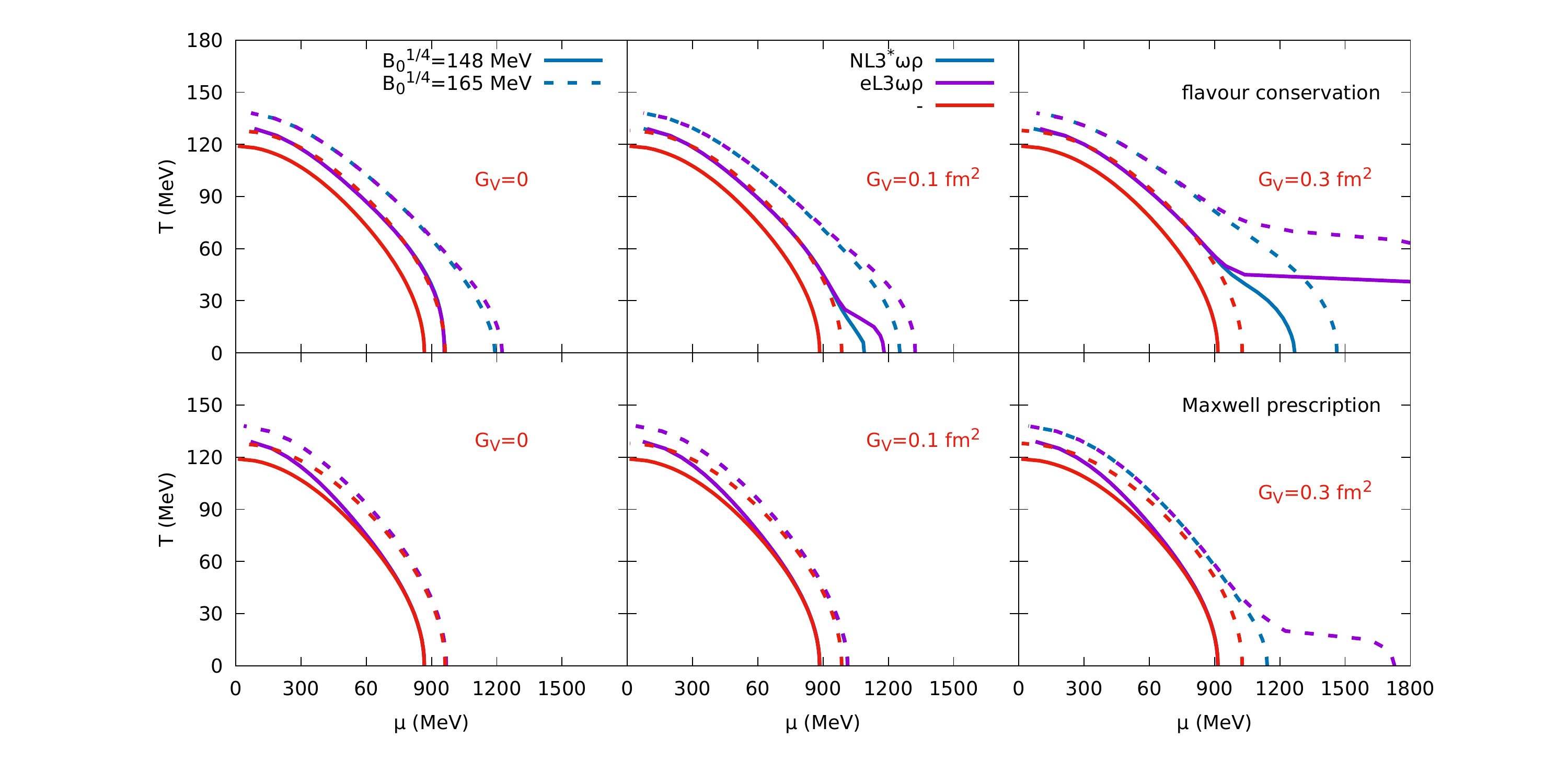}
\caption{Phase diagrams for strange stellar matter considering the NL3$^*\omega \rho$ and eL3$\omega \rho$ parametrizations for the hadronic matter and different MIT bag based models for two temperature dependent bag $B(T)$ values for the quark matter. At the top we use the prescription of the flavour conservation and at the bottom the of Maxwell.}  \label{fig:stellar_matter}
\end{centering}
\end{figure*}

In the following we present in Fig.~\ref{fig:stellar_matter} the phase diagrams for stellar matter and compare the results considering flavour conservation (top) and the Maxwell prescription (bottom). The blue lines are the ones where we use the NL3$^*\omega \rho$ parametrization, the purple ones are for the eL3$\omega \rho$ parametrization and the red ones we use to display, again, also the results obtained using only the MIT based models where, in this case, the quark matter is always $\beta$-stable. Furthermore, the solid lines stand for the results using a bag pressure value $B_0^{1/4}=148$~MeV and the dashed lines for $B_0^{1/4}=165$~MeV. The value of G$_V$ increases from 0 to 0.3 fm$^2$ from left to right.

As expected, the general behavior is the same as observed in Fig.~\ref{fig:symmetric_matter} for symmetric matter. For both prescriptions, only the bag pressure value dictates the maximum temperature obtained for $\mu \rightarrow 0$. At low temperatures, the maximum chemical potential depends largely on the bag pressure value, but also on the inclusion or not of the vector field to the bag model and the parametrization chosen for the QHD model.

As done above, we analyse first the region of low chemical potentials. At this region we have $T_{max}=129$~MeV for $B_0^{1/4}=148$~MeV and $T_{max}=138$~MeV for $B_0^{1/4}=165$~MeV no matter the QHD parametrization, if we add the vector channel to the MIT model or not or even if we use the prescription of Maxwell or of flavour conservation. Which is different from what happened in Fig.~\ref{fig:symmetric_matter}, where the results for $T_{max}$ are more or less the same, i.e. around $T_{max}=168$~MeV, no matter the value of $B_0$, but that is because we used the symmetric matter to chose the values of $T_0$ in Eq.~\ref{eq:B(T)} exactly so that the maximum temperatures where around this value
. Here, as we are not able to reach the same $T_{max}$, the maximum values of $B(T)$ are different for each value of $B_0$, which, in turn, means that the maximum temperatures here are also different and depend on the value of $B_0$. Also differently from what happened with symmetric matter, here the results for $T_{max}$ do not match the results obtained via MIT based model alone. This is so, for both prescriptions, because the crossing of the curves $\mu \times P$ occurs at pressures that are not so close to zero as in the symmetric matter case. At the same time, these pressures are low enough so that they fall in a region where the hadronic EoSs for stellar matter are stiffer than the ones for symmetric matter, and so we have again that a softer hadronic EoS favors the hadron phase.

Now we analyse the region at low temperatures. Here, as for the symmetric matter, the results are different for each combination of parametrizations and they also change depending on the prescription used, either flavour conservation or Maxwell construction. The maximum chemical potentials obtained for each combination of parametrizations are presented in Table~\ref{tab:pot_max_est}. 

\begin{widetext}
\begin{center}
\begin{table}[]
\caption{Chemical potentials ($\mu_{max}$) at $T=0$ for all combinations of parametrizations considering strange \textbf{stellar} matter. The first line corresponds to the results using the MIT based bag models only, where the phase transition criterion is just the value of the chemical potential where the pressure goes to zero, as done in \cite{lopes2021modified_pt2}. All values are given in MeV.}
\begin{tabular}{p{1.5cm}|p{2.5cm}|p{2.5cm}|p{2.5cm}|p{2.5cm}|p{2.5cm}|p{2.5cm}}
\toprule[0.6pt]
 Models  & $B_0^{1/4}=148$~MeV  & $B_0^{1/4}=148$~MeV + $G_V=0.1$~fm$^2$  & $B_0^{1/4}=148$~MeV + $G_V=0.3$~fm$^2$ & $B_0^{1/4}=165$~MeV  &   $B_0^{1/4}=165$~MeV + $G_V=0.1$~fm$^2$    &   $B_0^{1/4}=165$~MeV + $G_V=0.3$~fm$^2$   \\ \midrule[1.5pt]
 - & $\mu_{max}=867$  & $\mu_{max}=884$  & $\mu_{max}=915$  & $\mu_{max}=962$ &   $\mu_{max}=986$ &   $\mu_{max}=1027$  \\\midrule[1.5pt]
 \multicolumn{7}{c}{flavour conservation}\\
\bottomrule[0.6pt]
 NL3$^*\omega \rho$ & $\mu_{max}=960$  & $\mu_{max}=1089$  & $\mu_{max}=1268$  & $\mu_{max}=1192$   &   $\mu_{max}=1253$    &   $\mu_{max}=1462$  \\
 eL3$\omega \rho$  & $\mu_{max}=959$  & $\mu_{max}=1181$  & no crossing  & $\mu_{max}=1225$    &   $\mu_{max}=1324$    &   no crossing \\\midrule[1.5pt]
 \multicolumn{7}{c}{Maxwell prescription}\\
\bottomrule[0.6pt]
NL3$^*\omega \rho$ & $\mu_{max}=867$  & $\mu_{max}=884$  & $\mu_{max}=915$  & $\mu_{max}=967$   &   $\mu_{max}=1012$    &   $\mu_{max}=1142$  \\
 eL3$\omega \rho$  & $\mu_{max}=867$  & $\mu_{max}=884$  &  $\mu_{max}=915$ & $\mu_{max}=967$    &   $\mu_{max}=1013$    & $\mu_{max}=1727$   \\\bottomrule[0.6pt]
\end{tabular}\label{tab:pot_max_est}
\end{table}
\end{center}
\end{widetext}

Let us first scrutinize the effect of the different MIT bag based model
parametrizations at this region of low temperatures. 
We notice that the effect of $G_V$ here, in stellar matter, causes smaller changes in the $\mu_{max}$, which allows us to increase the values up to 0.3 fm$^2$. Nevertheless the qualitative results are the same: the higher the values for $G_V$ the higher are the results for $\mu_{max}$. We can still compare the results between symmetric matter and stellar matter when using the same parametrizations, namely, when $G_V=0$ and $G_V=0.1$~fm$^2$. In doing so we see very clearly that there is a reduction in the values of chemical potentials at $T=0$ here, being this reduction even greater when we use the Maxwell prescription. This reduction also grows with the value of $G_V$ and $B_0^{1/4}$ and is also greater when using the eL3$\omega \rho$ than when using the NL3$^*\omega \rho$ parametrization. 

Being that the quark EoSs for symmetric matter are always stiffer than the ones for stellar matter and that the ones for stellar matter resulting from the flavour conservation prescription are stiffer that the $\beta$-stable and charge neutral ones, in general, stiffer quark EoSs favor the hadron phase. The same is true when we compare the results for different values of $G_V$, as higher values of $G_V$ result in higher values for $\mu_{max}$ and also increase the stiffness of the EoS, the same behaviour found in symmetric matter. The increase of the bag pressure value $B_0^{1/4}$ also results in higher values of $\mu_{max}$, but it softens the EoS, as we already pointed out in \cite{biesdorf2022qcd} for symmetric matter. So, to summarize, stiffer quark EoSs always favor the hadron phase, unless we change the value of the Bag constant $B_0^{1/4}$ and then the opposite is true, a softer quark EoS favors the hadron phase.

An interesting aspect to analyse when comparing the results at low temperatures obtained from the two different prescriptions is their behavior in relation to the results obtained via MIT bag based model alone. As it is very clear from Fig.~\ref{fig:stellar_matter} and Table~\ref{tab:pot_max_est}, when we use the prescription of flavour conservation the results are quite different, whereas when using the Maxwell prescription the diagrams generally converge to the same $\mu_{max}$s. First we have to point out that the diagrams obtained via MIT bag based model alone (red lines) presented at the top and at the bottom of Fig.~\ref{fig:stellar_matter} are the same, i.e., of $\beta$-stable and charge neutral quark and lepton matter. Now, when we use a QHD+MIT models, with the flavour conservation the hadronic phase completely determines the quark phase through the bond given by Eq.~\ref{eq:charge_conser}. The quark matter at the phase transition point is not yet $\beta$-stable, and, adding to that, in this case, we have $P_0$s that are not so close to zero, so, it is not surprising that the results for this prescription are different from the ones obtain only with the MIT based model. When using the Maxwell prescription, however, we have a quark matter at the phase transition point that is $\beta$-stable, and, when the the Gibbs' conditions at this region are met at very low pressures ($P_0 \leq 5$~MeV/fm$^3$), we get $\mu_{max}$s very similar to the ones obtained via MIT alone. When the $P_0$s start to get higher, also the $\mu_{max}$s grow. 

As for the effects of the different QHD model parametrizations at this region of low temperatures, we can see that they are not so preponderant here as for the symmetric matter and are the smallest when using the Maxwell prescription. But, whenever there is a difference, the $\mu_{max}$s for the eL3$\omega \rho$ are the highest and in some cases, we were not able to find crossings at $T=0$, as for example, the case with $G_V=0.3$~fm$^2$ and this QHD parametrization. The reason for this behaviour is the fact that the onset of the hyperons softens the EoS in the hadronic phase, while the $G_V$ monotonically stiffens the EoS of the quark phase. At very high density we expect that the quark phase becomes non-interacting, therefore the contribution of the Dirac sea can no longer be neglected.

To finalize the analysis of the results in this region of low temperatures we point out that we obtain results for $\mu_{max}$ within a range of 1000 MeV from the $\mu=1200$~MeV indicated in \cite{ayriyan2018robustness} for $\beta$-stable matter and various combinations of parametrizations when using the prescription of flavour conservation, namely, for NL3$^*\omega \rho$ in combination with $B_0^{1/4}=148$ + $G_V=0.3$~fm$^2$, $B_0^{1/4}=165$ + $G_V=0$ and $B_0^{1/4}=165$ + $G_V=0.1$~fm$^2$ and for eL3$\omega \rho$ in combination with $B_0^{1/4}=148$ + $G_V=0.1$~fm$^2$ and $B_0^{1/4}=165$ + $G_V=0$, but only for one combination of parametrizations when using the Maxwell prescription, namely, for NL3$^*\omega \rho$ in combination with $B_0^{1/4}=165$ + $G_V=0.3$~fm$^2$. So, considering the MIT parametrizations, only two, the ones with $B_0^{1/4}=148$, are within the stability window and are able to describe stable strange matter \cite{lopes2021modified}. However, in order to construct hybrid stars, parametrizations outside the stability window are preferable to avoid that once the phase transition starts, the entire star converts to a strange quark star. In \cite{lopes2022hypermassive} we constructed hybrid stars using the original L3$\omega \rho$ parametrization (a slight different paramaetrization than the one used here) for the QHD model and the MIT based model. 

We could improve the results at this region by adding also a quartic term to the MIT vector model, as done in \cite{lopes2021modified,lopes2021modified_pt2}. Using the very small value of $b_4=2$ we are able to obtain crossings at T=0 for the two cases pointed out in table~\ref{tab:pot_max_est} where we did not obtain results without this term, although the values for the $\mu_{max}$, and also the pressure, would still be very high. Choosing higher values for $b_4$ we can adjust the values of $\mu_{max}$ without altering the values of $T_{max}$ and, the higher the value of $G_V$ the higher is the affect of this term, as already shown in \cite{lopes2021modified_pt2}. 



\subsection{Testing different hypotheses}

\begin{figure}[ht] 
\begin{centering}
\includegraphics[angle=0,width=0.5\textwidth]{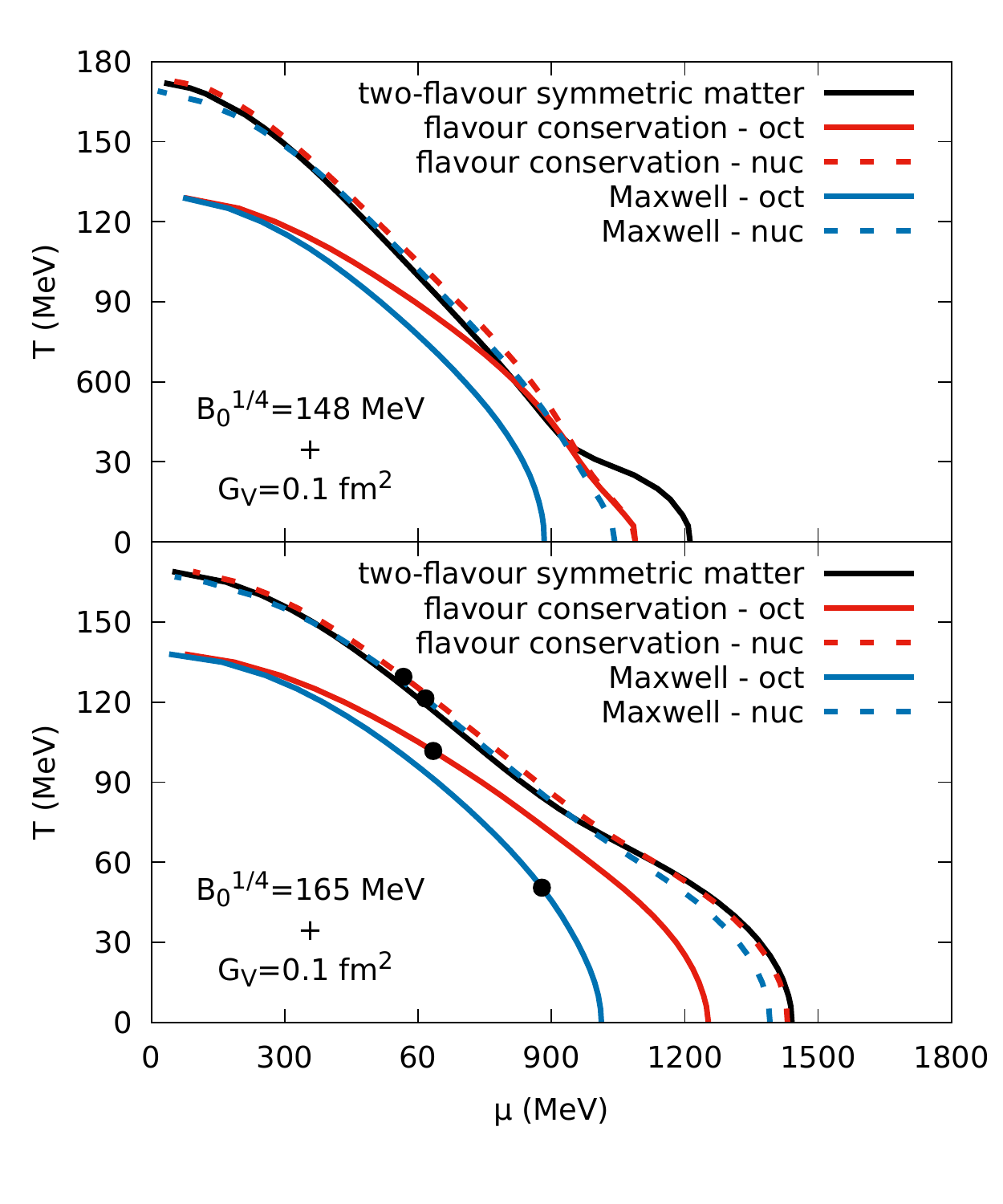}
\caption{Phase diagrams for all different matter hypothesis and prescriptions for phase transition used in this work considering the NL3$^* \omega \rho$ parametrization for the hadronic phase and $B_0^{1/4}=148$~MeV + $G_V=0.1$~fm$^2$ (top) and $B_0^{1/4}=165$~MeV + $G_V=0.1$~fm$^2$ (bottom) for the quark phase. We also include the results for stellar matter with no strange matter (dashed lines). The points presented on the bottom diagrams correspond to the ones where the latent heat $L|_S$ goes to zero (see section~\ref{sec:latent_heat}) and may correspond to the critical end points. } \label{fig:all_in_one}
\end{centering}
\end{figure}

In this section we take a better look at the differences occurring due to the different matter hypothesis and prescription imposed along this work. In order to do so we choose only two different combination of parametrizations and plot all results in  Fig.~\ref{fig:all_in_one}. We  also include the results for stellar matter with no strange matter, i.e., with no hyperons and strange quarks (dashed lines).

As already pointed out above and is more evident here, the maximum temperature for strange 
stellar matter depends only on the bag pressure value and is the same for both prescriptions, the one with flavour conservation (solid red line) and the one with Maxwell construction (solid blue line). It's also more evident here that at low temperatures the prescription of flavour conservation favors the hadronic phase. What is completely new here, though, are the diagrams for stellar matter with no hyperons in the hadronic matter and no strange quark in the quark matter. We chose only to present those results for this two combination of parametrizations because its general behavior follows the ones analysed until now, i.e., the different parametrizations have the same affect to this matter as for the previous cases.

As could be expected, the results for the diagrams for stellar matter without strangeness are very close to the ones for two-flavour symmetric matter. This because the contribution to the chemical potential coming from the leptons, only present in the stellar matter, is very small.

At very low temperatures we have obtained an interesting feature for the flavour conservation prescription. For the NL3$^* \omega \rho$ + $B_0^{1/4}=148$~MeV + $G_V=0.1$~fm$^2$ parametrization, the phase transition for the case where the strange matter is allowed to appear occurs at chemical potentials where there is no hyperon onset yet, so, at this region, the results for stellar matter with (solid red line) and without (dashed red lines) strange matter are the same. And at higher temperatures, where the hyperons start to appear at the point of the phase transition, the two lines start to differentiate and the line without strange matter approximates itself from the line of two-flavour symmetric matter. For the other parametrization, i.e., NL3$^* \omega \rho$ + $B_0^{1/4}=165$~MeV + $G_V=0.1$~fm$^2$, however, the phase transition for strange matter occurs at chemical potentials where there are already hyperons present and so, the results for stellar matter with and without strange matter are different throughout the entire phase diagram. As for the Maxwell prescription, despite the phase transition for strange matter at low temperatures for both parametrizations occurring at chemical potentials where there are no hyperons, the results for strange matter (solid blue lines) and no strange matter (dashed blue lines) are always very different. This happens because for strange quark stellar matter the strange quarks are always present, as can be seen in \cite{lopes2021modified}.

In Table \ref{tab:pot_max_est-ns} we present the results for the maximum chemical potentials at the phase transition for stellar matter without strange matter. There we also include the results when using only the MIT based model and, as can be noted, its results are only similar to the ones when combining models when we use $B_0^{1/4}=148$~MeV and $G_V=0$. This because in this cases $P_0 \rightarrow 0$ and for all the other parametrizations the phase transition happens at increasing pressures. Yet about the results for stellar matter without strange matter using only the MIT based model, its $T_{max}$s reach around only T=140 MeV, i.e., very below the ones obtained for the same matter but combining models. That's because when combining models the crossing of the $\mu \times P$ occurs at pressures considerably high, and higher than for strange matter.

\begin{widetext}
\begin{center}
\begin{table}[]
\caption{Chemical potentials ($\mu_{max}$) at $T=0$ for all combinations of parametrizations considering \textbf{stellar} matter \textbf{without strange matter}. All values are given in MeV }
\begin{tabular}{p{1.5cm}|p{2.5cm}|p{2.5cm}|p{2.5cm}|p{2.5cm}|p{2.5cm}|p{2.5cm}}
\toprule[0.6pt]
 Models  & $B_0^{1/4}=148$~MeV  & $B_0^{1/4}=148$~MeV + $G_V=0.1$~fm$^2$  & $B_0^{1/4}=148$~MeV + $G_V=0.3$~fm$^2$ & $B_0^{1/4}=165$~MeV  &   $B_0^{1/4}=165$~MeV + $G_V=0.1$~fm$^2$    &   $B_0^{1/4}=165$~MeV + $G_V=0.3$~fm$^2$   \\ \midrule[1.5pt]
 - & $\mu_{max}=953$  & $\mu_{max}=976$  & $\mu_{max}=1017$  & $\mu_{max}=1063$ &   $\mu_{max}=1094$ &   $\mu_{max}=1149$  \\\midrule[1.5pt]
 \multicolumn{7}{c}{flavour conservation}\\
\bottomrule[0.6pt]
 NL3$^*\omega \rho$ & $\mu_{max}=960$  & $\mu_{max}=1089$  & $\mu_{max}=1750$  & $\mu_{max}=1259$   &   $\mu_{max}=1431$    &   $\mu_{max}=1947$  \\
 eL3$\omega \rho$  & $\mu_{max}=959$  & $\mu_{max}=1568$  & no crossing  & $\mu_{max}=1405$    &   $\mu_{max}=1751$    &   no crossing \\\midrule[1.5pt]
 \multicolumn{7}{c}{Maxwell prescription}\\
\bottomrule[0.6pt]
NL3$^*\omega \rho$ & $\mu_{max}=955$  & $\mu_{max}=1042$  & $\mu_{max}=1671$  & $\mu_{max}=1229$   &   $\mu_{max}=1391$    &   $\mu_{max}=1886$  \\
 eL3$\omega \rho$  & $\mu_{max}=957$  & $\mu_{max}=1450$  &  no crossing & $\mu_{max}=1350$    &   $\mu_{max}=1681$    & no crossing   \\\bottomrule[0.6pt]
\end{tabular}\label{tab:pot_max_est-ns}
\end{table}
\end{center}
\end{widetext}

\section{Latent Heat and Latent Energy}\label{sec:latent_heat}

The latent heat is an important quantity in the study of phase transitions and we investigate two different expressions found in the literature.

We  first take a look at the relativistic latent heat $L|_\epsilon$, as given 
in ~\cite{agasian2008quark,lope2022maximum} 
\begin{equation}\label{eq:energia_latente}
    L|_\epsilon=P^H \frac{\epsilon^Q-\epsilon^H}{\epsilon^Q \epsilon^H},
\end{equation}
where $\epsilon^Q$ and $\epsilon^H$ are the energy densities at the point of the phase transition for the quark and hadronic matter, respectively. We call it latent energy next, since it yields results that are not zero at zero temperature. Notice that this quantity has no dimension. According to \cite{agasian2008quark}, when the latent energy becomes zero, the critical end point, where the first order phase transition turns into a crossover, is reached.

\begin{figure}[ht] 
\begin{centering}
\includegraphics[angle=0,width=0.5\textwidth]{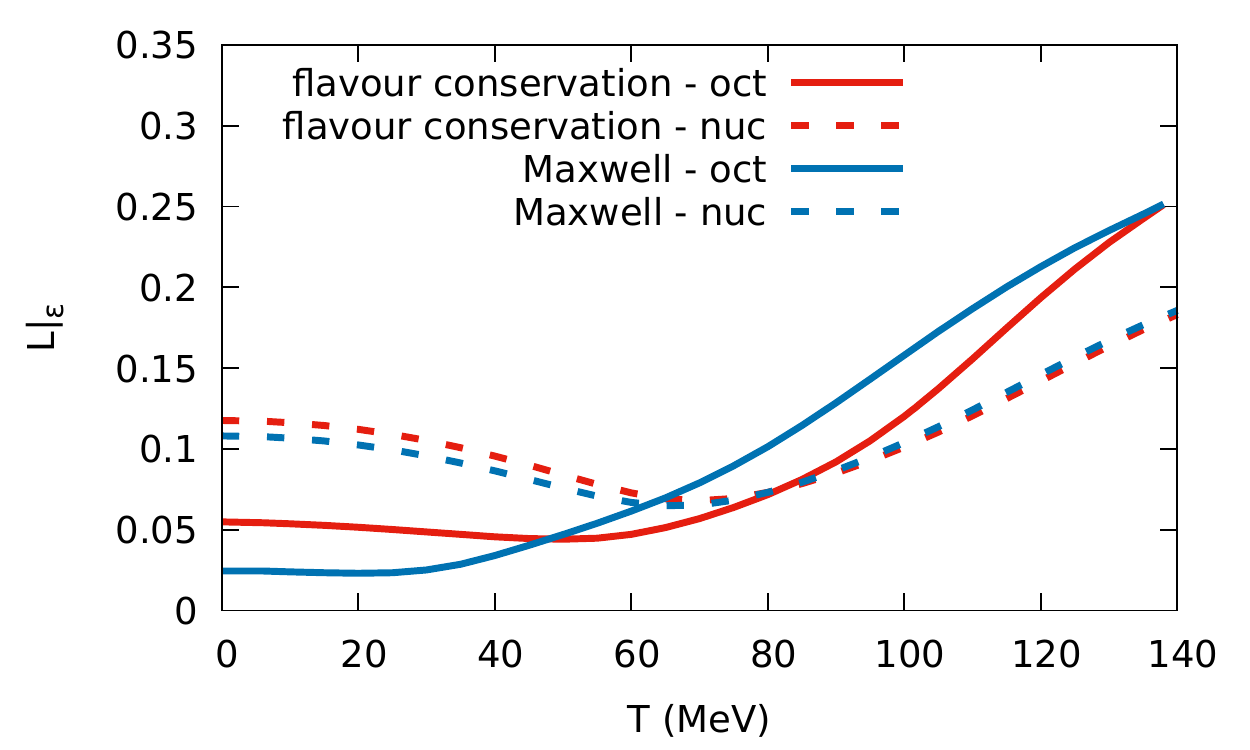}
\caption{Results for latent energy for different matter hypothesis and prescriptions for phase transition used in this work considering the NL3$^* \omega \rho$ parametrization for the hadronic phase and $B_0^{1/4}=165$~MeV + $G_V=0.1$~fm$^2$ for the quark phase.}\label{fig:energia_latente}
\end{centering}
\end{figure}

We also analyse the usual expression found in textbooks and investigated in \cite{roark2019hyperons} as well:
\begin{equation}\label{eq:calor_latente}
    L|_S= \left( S^Q - S^H \right) T,
\end{equation}
where $S^Q$ and $S^H$ are the entropy densities,  defined in Eq.~\ref{eq:entropia}, at the point of the phase transition for the quark and hadronic matter, respectively.

\begin{align}\label{eq:entropia}
    S&=- \gamma \sum_i \int \frac{d^3k}{(2 \pi)^3} \Bigg[f_{i+} \ln \left(\frac{f_{i+}}{1-f_{i+}} \right) + \\ \nonumber 
    & + \ln (1-f_{i+}) + f_{i-} \ln \left(\frac{f_{i-}}{1-f_{i-}} \right) + \ln (1-f_{i-}) \Bigg],
\end{align}
where $\gamma$ is degeneracy of spin (and color in the case of the quarks) and $f_{i \pm }$ are the Fermi-Dirac distribution functions of the particles and anti-particles, respectively.

\begin{figure}[ht] 
\begin{centering}
\includegraphics[angle=0,width=0.5\textwidth]{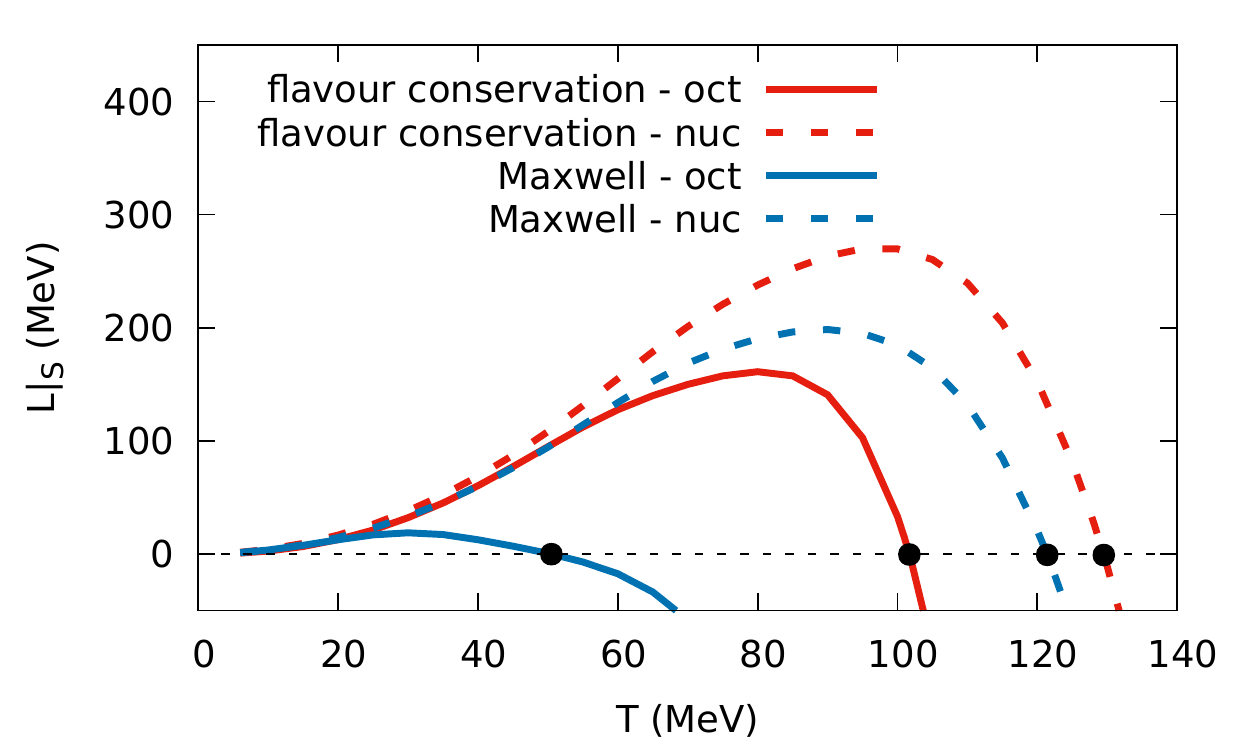}
\caption{Results for latent heat for different matter hypothesis and prescriptions for phase transition used in this work considering the NL3$^* \omega \rho$ parametrization for the hadronic phase and $B_0^{1/4}=165$~MeV + $G_V=0.1$~fm$^2$ for the quark phase.}\label{fig:calor_latente}
\end{centering}
\end{figure}

In Figs.~\ref{fig:energia_latente} and ~\ref{fig:calor_latente} we present the results obtained with the latent energy and latent heat, respectively as a function of the temperature at the point of the hadron-quark phase transition. We choose only one parametrization since they yield qualitatively 
similar results. One can see that our results for the latent energy at zero temperature are of the same order of magnitude as the ones obtained in \cite{pelicer2022phase}, but never reach the values shown in \cite{lope2022maximum} and are always positive. 
As far as the latent heat is concerned, the behaviour we find is more similar to the result shown in \cite{roark2019hyperons} and we do find a maximum point, but at much lower temperatures. Our curves then cross the zero value, which, according to \cite{agasian2008quark} is an indication of the end of the first order phase transition. Assuming this interpretation is correct, our curves depicted in the QCD phase diagrams are only valid until the temperatures where $L|_S$ is zero and the corresponding values are shown as dots also in Fig.\ref{fig:all_in_one}.

\section{Conclusions}

In this work, we sought to reproduce the QCD phase diagram using two effective models. For the description of quark matter, a simple relativistic model, the MIT bag model and a modification thereof, were used, the latter being new to this kind of procedure. And for the description of hadronic matter, a model of quantum hadrodynamics, the Walecka Model with non-linear terms.

The Gibbs' conditions were used to establish the crossing points of the pressure as a function of the chemical potentials obtained in both phases.

Some restrictions were imposed when choosing the models mentioned above. We used temperature-dependent Bag B(T) values with the same parametrizations already used in our previous work \cite{lopes2021modified_pt2}, with one set of constants within the limits the stability window, i.e., $B_0^{1/4}=148$~MeV and another set outside this window , i.e., $B_0^{1/4}=165$~MeV, both sets with $G_V$ varying from zero to $G_V=0.3$~fm$^2$. As for the QHD based model, we used the NL3$^*\omega \rho$ and eL3$\omega \rho$ parametrizations, which satisfy several nuclear and astrophysical properties.

We first obtained the phase diagrams for two-flavour symmetric matter. As the parametrizations for the B(T) were adjusted so that the $T_{max}$ would satisfy the restrain imposed by the LQCD and freeze-out results, the $T_{max}$s are all around 168 MeV, but only the results for $B_0^{1/4}=165$~MeV fit the Cleymans line entirely inside the confined (hadron) phase. The results for low temperatures depend on all parametrizations, as already concluded in \cite{biesdorf2022qcd}. The influences of the different MIT based models follow the conclusions of \cite{lopes2021modified_pt2}, but here the results are more sensible to the different values of $G_V$s. Only a few combination of parametrizations resulted in maximum chemical potentials $\mu_{max}$ within the range $1050 \leq \mu \leq 1400$~MeV. Ultimately only NL3$^*\omega \rho$ + $B_0^{1/4}=165$~MeV and NL3$^*\omega \rho$ + $B_0^{1/4}=165$~MeV + $G_V=0.05$~fm$^2$ satisfy all restrains we imposed.

We also calculated the phase diagrams for stellar matter using two different prescriptions, one where we impose flavour conservation at the point of the phase transition, a prescription never applied to finite temperatures before, and the other where both phases are $\beta$-stable and charge neutral, being this last one the Maxwell prescription. Between the two prescriptions, the flavour conservation results in higher chemical potentials at low temperatures, but for both prescriptions the $T_{max}$s are the same and depend only on the value of $B_0$. Comparing this results to the symmetric matter, we conclude that the stellar matter favors the quark phase, being that the $T_{max}$s and $\mu_{max}$s are higher for symmetric matter. 

In all cases a higher Bag pressure and $G_V$ value increases the value of the chemical potentials at low temperatures, but this increase is lower for the stellar matter in the flavour conservation prescription and even lower in the Maxwell prescription than in the two-flavour symmetric matter. As for the QHD model parametrization, the eL3$\omega \rho$ always favors the hadron phase when compared to the NL3$^*\omega \rho$ one.

We also discussed the results for stellar matter without strange matter end concluded that those results are very similar to the ones for two-flavour symmetric matter.

Finally, at the last section we discussed briefly the results for latent heat and showed that for stellar matter with strange matter it always grows with the temperature. For two-flavour symmetric matter and stellar matter without strange matter, though, the behavior is more of a sinusoid.

\hspace{5cm}

{\bf Declarations}

\textbf{Funding:} this work is a part of the project INCT-FNA Proc. No. 464898/2014-5. D.P.M. was partially supported by Conselho Nacional de Desenvolvimento Científico e Tecnológico (CNPq/Brazil) under grant 303490-2021-7  and  C.B. acknowledges a doctorate scholarship from Coordenação de Aperfeiçoamento de Pessoal do Ensino Superior (Capes/Brazil). 

\textbf{Conflicts of interest/Competing interests:} the authors have no conflicts of interest to declare that are relevant to the content of this article.

\vspace{0.2cm}
{\bf Acknowledgments} 
C.B. thanks fruitful discussions with Mateus R. Pelicer.



\begin{widetext}
   \section*{References} 
\end{widetext}

\bibliography{references}




\end{document}